\documentclass[conference]{IEEEtran}
\IEEEoverridecommandlockouts

\usepackage{cite}
\usepackage{amsmath,amssymb,amsfonts}
\usepackage{algorithmic}
\usepackage{graphicx}
\usepackage{textcomp}
\usepackage{xcolor}
\usepackage{empheq}
\usepackage{hyperref}
\usepackage{algorithm,algorithmic}
\usepackage{multirow}
\usepackage{booktabs}
\usepackage{siunitx}
\usepackage{tabularx}
\newcommand{\R}{\ensuremath{\mathbb{R}}}
\DeclareMathOperator\artanh{artanh}
\def\BibTeX{{\rm B\kern-.05em{\sc i\kern-.025em b}\kern-.08em
		T\kern-.1667em\lower.7ex\hbox{E}\kern-.125emX}}
\floatname{algorithm}{Algorithm}
\newcommand{\KwIn}[1]{\textbf{Input:} #1}
\newcommand{\KwOut}[1]{\textbf{Output:} #1}
\newcommand{\sumlim}[2]{\ensuremath{\sum\limits_{#1}^{#2}}}
\makeatletter
\renewcommand{\vec}[1]{%
	\ifcat\relax\noexpand#1%
	\ensuremath{\boldsymbol{\lowercase{#1}}}%
	\else
	\ensuremath{\mathbf{\lowercase{#1}}}%
	\fi
}
\makeatother
\begin{document}
	
	\title{Performance Modeling for Correlation-based Neural Decoding of Auditory Attention to Speech\\
		\thanks{This work was supported by a junior postdoctoral fellowship fundamental research of the FWO (for S. Geirnaert, No. 1242524N), FWO project nr. G081722N, the Flemish Government (AI Research program), Internal Funds KU Leuven (project IDN/23/006), and the European Research Council (ERC) under the European Union’s research and innovation programme (No 101138304). Views and opinions expressed are however those of the author(s) only and do not necessarily reflect those of the European Union or any of the granting authorities. Neither the European Union nor the granting authorities can be held responsible for them. S. Geirnaert, J. Vanthornhout, T. Francart, and A. Bertrand are affiliated to Leuven.AI - KU Leuven institute for AI, Leuven, Belgium.}
	}
	
	\author{\IEEEauthorblockN{Simon Geirnaert}
		\IEEEauthorblockA{\textit{Dept. of Electrical Engineering} \\
			\textit{KU Leuven}\\
			Leuven, Belgium \\
			{\small simon.geirnaert@kuleuven.be}}
		\and 
		\IEEEauthorblockN{Jonas Vanthornhout}
		\IEEEauthorblockA{\textit{Dept. of Neurosciences} \\
			\textit{KU Leuven}\\
			Leuven, Belgium \\
			{\small jonas.vanthornhout@kuleuven.be}}
		\and
		\IEEEauthorblockN{Tom Francart}
		\IEEEauthorblockA{\textit{Dept. of Neurosciences}\\
			\textit{KU Leuven}\\
			Leuven, Belgium \\
			{\small tom.francart@kuleuven.be}}
		\and
		\IEEEauthorblockN{Alexander Bertrand}
		\IEEEauthorblockA{\textit{Dept. of Electrical Engineering} \\
			\textit{KU Leuven}\\
			Leuven, Belgium \\
			{\small alexander.bertrand@kuleuven.be}}
	}
	
	\maketitle
	
	\begin{abstract}
		Correlation-based auditory attention decoding (AAD) algorithms exploit neural tracking mechanisms to determine listener attention among competing speech sources via, e.g., electroencephalography signals. The correlation coefficients between the decoded neural responses and encoded speech stimuli of the different speakers then serve as AAD decision variables. A critical trade-off exists between the temporal resolution (the decision window length used to compute these correlations) and the AAD accuracy. This trade-off is typically characterized by evaluating AAD accuracy across multiple window lengths, leading to the performance curve. We propose a novel method to model this trade-off curve using labeled correlations from only a single decision window length. Our approach models the (un)attended correlations with a normal distribution after applying the Fisher transformation, enabling accurate AAD accuracy prediction across different window lengths. We validate the method on two distinct AAD implementations: a linear decoder and the non-linear VLAAI deep neural network, evaluated on separate datasets. Results show consistently low modeling errors of approximately 2 percent points, with 94\% of true accuracies falling within estimated 95\%-confidence intervals. The proposed method enables efficient performance curve modeling without extensive multi-window length evaluation, facilitating practical applications in, e.g., performance tracking in neuro-steered hearing devices to continuously adapt the system parameters over time.
	\end{abstract}
	
	
	\section{Introduction}
	\label{sec:intro}
	\noindent
	Current hearing devices often struggle in cocktail party scenarios with multiple competing conversations. Selective auditory attention decoding (AAD) addresses this by determining the attended speech signal from neural recordings such as electroencephalography (EEG)~\cite{mesgarani2012selective,osullivan2014attentional}. AAD has, therefore, important applications in so-called neuro-steered hearing devices, allowing them to automatically steer beamformers or source separation algorithms towards the attended conversation~\cite{geirnaert2021eegBased}.
	
\begin{figure*}
		\centering
		\includegraphics[width=0.775\textwidth]{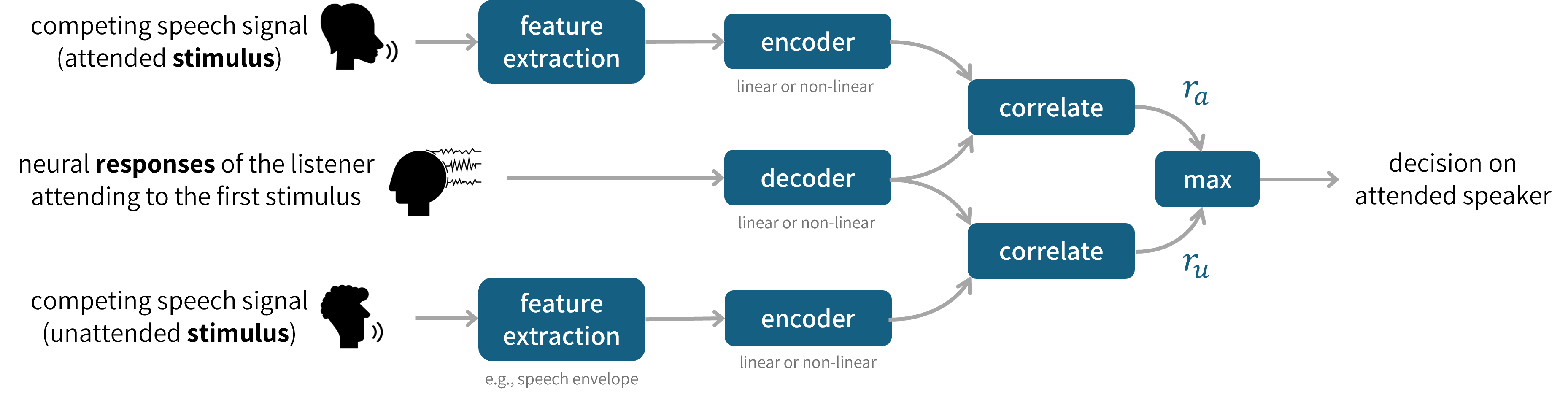}
		\caption{In this paper, we start from correlation-based AAD algorithms that use a stimulus encoder and/or neural decoder of which the outputs are correlated to determine the attended speaker.}
		\label{fig:decoding-pipeline}
	\end{figure*}

	Various AAD algorithms leverage the neural tracking phenomenon~\cite{osullivan2014attentional,geirnaert2021eegBased}, referring to the synchronization of the listener's neural signals with features of the attended speech signal~\cite{mesgarani2012selective}. As illustrated in Fig.~\ref{fig:decoding-pipeline}, these algorithms employ a (linear or non-linear) neural decoder and/or stimulus encoder to generate representations that correlate well. The strength of this relationship is then quantified using the Pearson correlation coefficient, which serves as a decision variable to determine attention to one of multiple speech sources.
	
	A critical parameter in AAD algorithms is the \emph{decision window length}. In neural tracking algorithms, this corresponds to the number of samples used to compute the correlation coefficient. Therefore, this parameter determines the time resolution at which attention decisions are made. However, there exists a fundamental trade-off: shorter windows improve the temporal resolution but typically reduce accuracy. This trade-off is captured in the algorithm's performance curve, showing AAD accuracy against decision window length~\cite{geirnaert2020interpretable}.
	
	Quantification of this performance curve is crucial for optimizing AAD-based gain control systems by determining the optimal decision window length. This is done, for example, based on the minimal expected switch duration~\cite{geirnaert2020interpretable}, which is a metric that uses the performance curve as an input. Currently, generating this performance curve requires evaluating the AAD algorithm across \emph{multiple} decision window lengths. However, this approach becomes impractical for efficiently tracking the performance curve in real time, e.g., to adapt gain control system parameters in a neuro-steered hearing device over time~\cite{geirnaert2022timeAdaptive}. 
	
	To address this, we propose a new method to model the complete performance curve of correlation-based AAD algorithms using (known or measured) correlation statistics for only \emph{one} decision window length. This enables continuous and efficient monitoring of the trade-off between accuracy and decision window length, without having to evaluate the model for all possible window lengths. Furthermore, the proposed technique provides a theoretical model to inform new correlation-based AAD algorithm design by translating target accuracy requirements into corresponding correlation coefficient targets, and vice versa. As this is an accuracy evaluation framework, we assume ground-truth attention labels are known, and leave the integration with unsupervised accuracy estimation as in L{\'o}pez-Gordo et al.~\cite{gordo2025unsupervised} for future work.
	
	The proposed method is explained in Section~\ref{sec:modeling}. Section~\ref{sec:experiment} describes the experiment setup, including the datasets and AAD algorithms used for validating the proposed method. Results and discussion are provided in Section~\ref{sec:results}.
	
	\section{Performance curve modeling based on one decision window length}
	\label{sec:modeling}
	\noindent
	
	\subsection{Correlation-based AAD algorithms}
	\label{sec:correlation-based-AAD}
	\noindent
	In this paper, we start from correlation-based AAD algorithms that use neural tracking, as shown in Fig.~\ref{fig:decoding-pipeline}. These algorithms employ both stimulus encoders and neural response (EEG) decoders, which can be implemented using either linear or non-linear approaches (see Section~\ref{sec:experiment} for specific examples).  
	
	For simplicity, we consider the case of two competing speakers, for which the AAD pipeline compares correlation coefficients between the decoded response and each encoded competing speech signal, identifying the highest correlation as the attended speaker. This approach is based on the fundamental assumption that the expected correlation with the attended speaker is higher than the expected correlation of the unattended speaker(s).
	
	\subsection{Modeling of correlations in the AAD decision system}
	\label{sec:modeling-correlations}
	\noindent
	In this paper, we evaluate the AAD accuracy assuming the ground-truth labels are known, as opposed to the unsupervised accuracy estimation technique in L{\'o}pez-Gordo et al.~\cite{gordo2025unsupervised}. For a given window length $w$ and sampling frequency $f_s$, the measured correlation coefficient $r_a$ over $N = wf_s$ samples between the (zero-mean) decoded response $\vec{x}\in\R^N$ and encoded attended speech signal $\vec{y}_a \in \R^N$ corresponds to:
	\begin{equation}
		\label{eq:corr}
		r_a = \frac{\sumlim{n=1}{N}x_ny_{a,n}}{\sqrt{\sumlim{n=1}{N}x_n^2}\sqrt{\sumlim{n=1}{N}y_{a,n}^2}},
	\end{equation}
	with $\rho_a = \mathbb{E}\!\left\{r_a\right\}$ the expected attended correlation (and similarly for the unattended speech $\vec{y}_u$, resulting in the (expected) unattended correlations $r_u$ and $\rho_u$).
	
	Previous work~\cite{geirnaert2025modeling} has shown that stimulus-response correlations in neural tracking such as $r_{a/u}$ can then be effectively modeled using a normal distribution after applying the Fisher transformation (i.e., using an inverse hyperbolic tangent)~\cite{fisher1921on}:
	\[
	z_{a/u} = \artanh\!\left(r_{a/u}\right) \sim \mathcal{N}\!\left(\mu_{a/u},\sigma_{a/u}^2\right).
	\]
	This modeling approach has been validated in the context of neural tracking to single speakers~\cite{geirnaert2025modeling} and is further supported by our experimental results in Section~\ref{sec:experiment}.
	
	Since the AAD decision rule compares per-window attended and unattended correlations ($r_a$ and $r_u$), and given that the inverse hyperbolic tangent is \emph{monotonically increasing}, we can use $\text{max}\!\left(z_a,z_u\right)$ instead of $\text{max}\!\left(r_a,r_u\right)$ in Fig.~\ref{fig:decoding-pipeline} to make the binary decision on attention. Under the assumption that $r_a$ and $r_u$ are uncorrelated (as empirically validated in L{\'o}pez-Gordo et al.~\cite{gordo2025unsupervised}), we can use $z_a-z_u \sim \mathcal{N}\!\left(\mu_a-\mu_u,\sigma_a^2+\sigma_u^2\right) \lessgtr 0$ as our decision variable. The AAD accuracy can then be computed using the cumulative distribution function (CDF) $F(\cdot)$ of the normal distribution, evaluated in zero~\cite{gordo2025unsupervised}:
	\begin{equation}
		\label{eq:aad-acc}
		\text{AAD accuracy} = 100\left(1-F(0;\mu_a-\mu_u,\sigma_a^2+\sigma_u^2)\right).
	\end{equation}
	
	To determine the parameters required for computing the AAD accuracy in~\eqref{eq:aad-acc}, we employ the first-order Hotelling series expansion to approximate the mean and variance of the Fisher-transformed variables $z_{a/u}$~\cite{hotelling1953new,fouladi2008fisher}:
	\[
	\mu_{a/u} \approx \artanh\!\left(\rho_{a/u}\right)+\frac{\rho_{a/u}}{2(N-1)}, \sigma_{a/u}^2 \approx \frac{1}{N-1},
	\]
	where $N = w f_s$ represents the number of samples used in the correlation computation in \eqref{eq:corr}, derived from the decision window length $w$ and sampling frequency $f_s$. With this, the parameters of the decision variable distribution $\mathcal{N}\!\left(\mu_a-\mu_u,\sigma_a^2+\sigma_u^2\right)$ can be approximated as:
	\begin{empheq}[left=\empheqlbrace]{align}
		\mu_a-\mu_u &\approx \artanh\!\left(\rho_{a}\right)-\artanh\!\left(\rho_{u}\right)+\frac{\rho_a-\rho_u}{2(N-1)},\label{eq:params-aad-dec-mu}\\
		\sigma_{a}^2+\sigma_u^2 &\approx \frac{2}{N-1}\label{eq:params-aad-dec-sigma}.
	\end{empheq}

	\subsection{Modeling the entire performance curve}
	\label{sec:modeling-performance-curve}
	\noindent
	A fundamental insight underpinning our method is that the mean of the measured correlation values $r_{a/u}$ (represented by $\rho_{a/u}$) remains constant across different decision window lengths, as demonstrated in~\cite{gordo2025unsupervised}. Only the variance of $r_{a/u}$ changes with window length $w$. Leveraging this property, we can derive the relationship between parameters at different window lengths. Let $w_1$ be the baseline window length for which we effectively compute the $r_{a/u}$ values and $w_2$ the target window length. By subtracting $\mu_a^{(1)}-\mu_u^{(1)}$ at baseline window length $w_1$ from $\mu_a^{(2)}-\mu_u^{(2)}$ at $w_2$ using \eqref{eq:params-aad-dec-mu}, we obtain:
	\begin{equation}
		\label{eq:correction-mu}
		\mu_a^{(2)}-\mu_u^{(2)} \approx \mu_a^{(1)}-\mu_u^{(1)} + \frac{(N_2-N_1)(\rho_a-\rho_u)}{2(N_2-1)(N_1-1)},
	\end{equation}
	where $N_1 = w_1 f_s$ and $N_2 = w_2 f_s$. Both $\mu_a^{(1)}-\mu_u^{(1)}$ and $\rho_{a/u}$ can be estimated from the available correlations $r_{a/u}$ at the initial window length $w_1$.
	
	Similarly, the variance at the target window length $w_2$ can be derived by dividing \eqref{eq:params-aad-dec-sigma} between the two window lengths:
	\begin{equation}
		\label{eq:correction-sigma}
		\sigma_a^{(2)^2}+\sigma_u^{(2)^2} \approx \left(\sigma_a^{(1)^2}+\sigma_u^{(1)^2}\right)\frac{N_1-1}{N_2-1}.
	\end{equation}
	This relationship confirms the intuition that larger window lengths lead to reduced variance in the decision variable, as more samples are available for correlation computation.
	
	
	Eq.~\eqref{eq:correction-mu} and \eqref{eq:correction-sigma} provide the necessary correction factors to extrapolate the parameters of the Fisher-transformed normal distribution from one decision window length to another. This enables modeling of the entire performance curve from correlations measured at a single window length. To quantify the uncertainty in the predictions of the accuracies, we employ the bias-corrected and accelerated bootstrapping method~\cite{diciccio1996bootstrap} to compute $95\%$-confidence intervals for the estimated AAD accuracy. $1000$ bootstrap samples are created. Algorithm~\ref{algo:acc-pred} presents a complete summary of our proposed method, and all MATLAB code can be found online~\cite{geirnaert2025aad}.
	
	\begin{algorithm}
		\caption{Accuracy prediction for performance curve modeling in correlation-based AAD algorithms}
		\label{algo:acc-pred}
		\KwIn{$M$ labeled correlation coefficients $\{r_{a/u}^{(m)}\}_{m=1}^M$ at decision window length $w_1$ resulting from a correlation-based AAD algorithm as in Fig.~\ref{fig:decoding-pipeline}, sampling frequency $f_s$, target decision window length $w_2$\\
			\KwOut{AAD accuracy at $w_2$}}
		\begin{algorithmic}[1]
			\STATE Compute $\rho_{a/u} = \underset{\text{over } m}{\text{mean}}(r_{a/u}^{(m)})$
			\STATE Apply the Fisher transformation $z_{a/u}^{(m)} = \artanh\!\left(r_{a/u}^{(m)}\right)$ and compute from the given correlations
			\begin{equation*}
				\begin{cases}
					\mu_a^{(1)}-\mu_u^{(1)} = \underset{\text{over } m}{\text{mean}}(z_a^{(m)}-z_u^{(m)}),\\ \sigma_a^{(1)^2}+\sigma_u^{(1)^2} = \underset{\text{over } m}{\text{variance}}(z_a^{(m)}-z_u^{(m)})
				\end{cases}
			\end{equation*}
			\STATE Extrapolate parameters to window length $w_2$, with $N_1 = w_1 f_s, N_2 = w_2 f_s$:
			\begin{equation*}
				\begin{cases}
					\mu_a^{(2)}-\mu_u^{(2)} = \mu_a^{(1)}-\mu_u^{(1)} + \frac{(N_2-N_1)(\rho_a-\rho_u)}{2(N_2-1)(N_1-1)},\\
					\sigma_a^{(2)^2}+\sigma_u^{(2)^2} = \left(\sigma_a^{(1)^2}+\sigma_u^{(1)^2}\right)\frac{N_1-1}{N_2-1}.
				\end{cases}
			\end{equation*}
			\STATE Predict AAD accuracy at $w_2$ using the CDF $F(\cdot)$ of the normal distribution at 0 with the extrapolated parameters:
			\[
			\text{accuracy} = 100\left(1-F(0;\mu_a^{(2)}-\mu_u^{(2)},\sigma_a^{(2)^2}+\sigma_u^{(2)^2})\right)
			\]
		\end{algorithmic}
	\end{algorithm}
	
	\section{Experiment setup}
	\label{sec:experiment}
	\noindent
	To demonstrate the broad applicability of our performance curve modeling method, we validate it using two variants of the correlation-based AAD pipeline shown in Fig.~\ref{fig:decoding-pipeline}: a linear decoder~\cite{osullivan2014attentional,geirnaert2021eegBased} and the non-linear VLAAI deep neural network~\cite{accou2023decoding}. Each variant is evaluated on a different dataset for generalization purposes.
	
	\subsection{Linear decoding on Das-2016 dataset}
	\label{sec:linear-mmse-das}
	\noindent
	The first setup uses the commonly used linear least-squares decoder (without encoder) to reconstruct the attended speech envelope in the $1-\SI{9}{\hertz}$ frequency range~\cite{osullivan2014attentional,geirnaert2021eegBased}. This decoder is a spatio-temporal filter that integrates EEG channels across time lags to minimize the squared error between the reconstructed and attended speech envelopes. We follow all preprocessing and implementation details from~\cite{geirnaert2022timeAdaptive}, with the sole modification of using a $\left[100,400\right]$ms integration window.
	
	This AAD algorithm is evaluated on the publicly available Das-2016 dataset, which contains EEG recordings from 16 participants attending to one of two competing speakers for $\SI{72}{\minute}$~\cite{biesmans2017auditory}. We compute the attended and unattended correlations using user-specific 5-fold cross-validation on $\SI{60}{\second}$ trials. Ground-truth AAD accuracies and (un)attended correlations are computed across decision window lengths of $\{60,30,20,10,5,1\}$s. To analyze the impact of the amount of estimation data, we vary the amount of data used ($\{2,5,10,30,60\}$ min) by randomly sampling from the available test correlations, which is repeated ten times per condition. This AAD algorithm and evaluation serves as the default experimental setup in Section~\ref{sec:results} unless specified otherwise.

	\subsection{VLAAI deep neural network on Fuglsang-2018 dataset}
	\label{sec:vlaai-fuglsang}
	\noindent
	The second independent validation employs the non-linear VLAAI deep neural network, which is pre-trained on a large single-speaker dataset to reconstruct the $1-\SI{32}{\hertz}$ speech envelope~\cite{accou2023decoding}. We apply this network as a user-independent AAD decoder to the publicly available Fuglsang-2018 dataset, containing EEG recordings from 18 participants attending to one of two competing talkers, with 60 trials of $\SI{50}{\second}$ per participant~\cite{fuglsang2017noise}. Data preprocessing follows~\cite{accou2023decoding}, except for the multi-channel Wiener filter which is replaced by regressing out the EOG from EEG channels for artifact removal. Ground-truth AAD accuracies and correlations are computed for decision window lengths of $\{50,25,10,5,1\}$s.
	
	\section{Results and discussion}
	\label{sec:results}
	\noindent
	We evaluate our performance modeling method to estimate the average performance curve (Section~\ref{sec:dataset-wide}), investigate the impact of the amount of estimation data (Section~\ref{sec:amount-of-estimation-data}), and show per-participant performance curve estimation (Section~\ref{sec:individual-est}). All experiments use the linear decoder on the Das-2016 dataset (Section~\ref{sec:linear-mmse-das}), while evaluation on the VLAAI setup is provided in Section~\ref{sec:validation-vlaai}.
	
	\subsection{Average performance curve estimation}
	\label{sec:dataset-wide}
\begin{figure}
		\centering
		\includegraphics[width=1\linewidth]{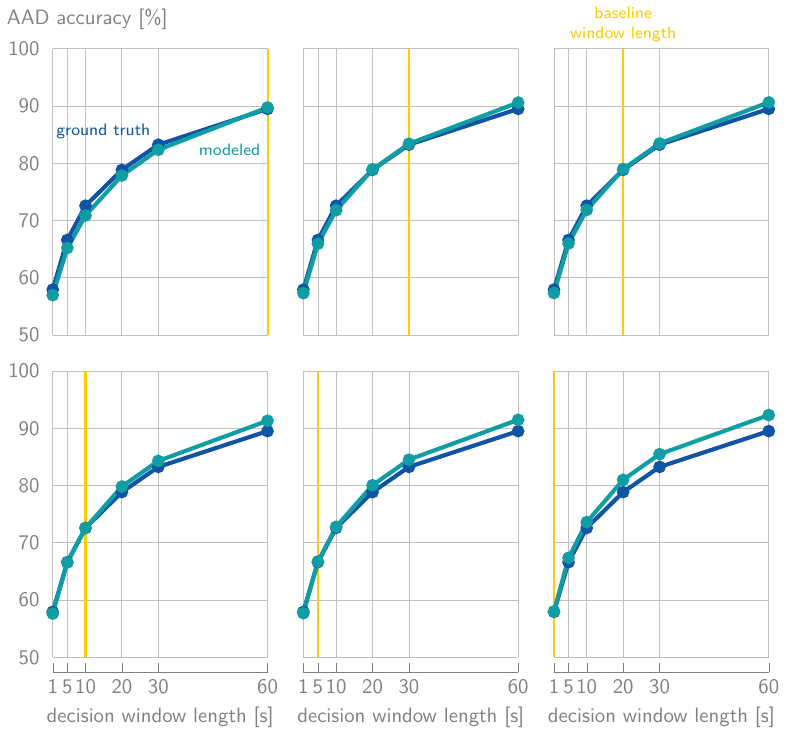}
		\caption{The modeled (predicted) vs. ground-truth average performance curves of the linear decoder on the Das-2016 dataset, starting from various baseline window lengths indicated in yellow and using $\SI{60}{\minute}$ of estimation data.}
		\label{fig:dataset-wide-wl}
	\end{figure}

	\noindent
	We first assess the method's ability to model performance curves across all participants. Fig.~\ref{fig:dataset-wide-wl} compares predicted and ground-truth average performance curves (by averaging the per-participant modeled performance curves) for different baseline decision window lengths, using $\SI{60}{\minute}$ of randomly sampled estimation data (ten repetitions).
	
	The first overall observation is that the modeled performance curves generally closely approximate the ground truth. As expected, optimal prediction accuracy is observed at the corresponding baseline window length, confirming the modeling of the AAD decision variable using a normal distribution after the Fisher transformation. Prediction error increases with the extrapolation distance between baseline and target window lengths. The baseline window length can, therefore, be tailored towards the range of window lengths of interest (for example, the shorter ones in the context of AAD gain control systems~\cite{geirnaert2020interpretable}). In general, a good choice is in the middle of the considered range of window lengths. Here, based on Fig.~\ref{fig:dataset-wide-wl}, we select $\SI{20}{\second}$ as the baseline for subsequent analyses.
	
	\subsection{Amount of estimation data}
	\label{sec:amount-of-estimation-data}
\begin{figure}
		\centering
		\includegraphics[width=0.775\linewidth]{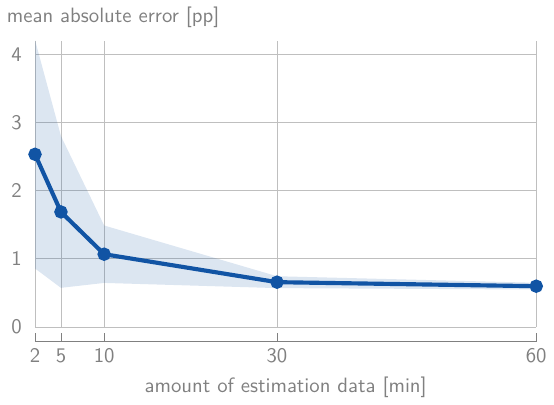}
		\caption{The mean ($\pm$ standard deviation) absolute error in percent point (pp) between the modeled and true AAD accuracy across the average performance curve of the linear decoder on the Das-2016 dataset in fucntion of the amount of estimation data.}
		\label{fig:amount-estimation-data}
	\end{figure}

	\noindent
	Using the $\SI{20}{\second}$ baseline window length, we analyze prediction performance across varying amounts ($\{2,5,10,30,60\}$ min) of estimation data. Per amount of estimation data, a corresponding number of random $\SI{20}{\second}$-window correlations are sampled, which is repeated ten times. Fig.~\ref{fig:amount-estimation-data} shows the mean absolute percent point (pp) error between modeled and true AAD accuracy across the average dataset-wide performance curve, with uncertainty bands indicating standard deviation across the ten repetitions.
	
	As expected, the prediction error and its variability increase when less estimation data are available, yet remains reasonably low ($\approx 2.5$ pp) even with only $\SI{2}{\minute}$ of estimation data. The prediction error stabilizes beyond $\SI{30}{\minute}$, which we adopt in subsequent analyses. In practical applications, however, the AAD accuracy and correlation distributions can change over time due to variations in the EEG signal quality, user performance, or when time-adaptive decoders are used~\cite{geirnaert2022timeAdaptive}. Such cases warrant a smaller amount of estimation data to better track the performance curve over time, at the cost of larger performance estimation errors as shown in Fig.~\ref{fig:amount-estimation-data}.
	
	\subsection{Individual performance curve estimation}
	\label{sec:individual-est}
	\noindent
	Table~\ref{tab:ind-perf} presents per-participant ground-truth and predicted accuracies, mean and standard deviation of the absolute error in pp, and the proportion of true accuracies within the estimated $95\%$-confidence interval, using $\SI{30}{\minute}$ of estimation data and $\SI{20}{\second}$ baseline window length, averaged across ten repetitions.

		The results clearly show that the accuracy prediction method closely approximates the ground-truth accuracy, even on a per-participant level, with a mean absolute error of $2.1$ pp (standard deviation $1.7$ pp). The estimated $95\%$-confidence intervals contain the true accuracy in $94.0\%$ of cases, validating our uncertainty quantification.
		
		\begin{table}[]
			\setlength{\tabcolsep}{1.5pt}
			\caption{The per-participant prediction results for the linear decoder on the Das-2016 dataset, starting from $\SI{20}{\second}$ baseline window length and using $\SI{30}{\minute}$ of estimation data. {\tiny CI = confidence interval}}
			\centering
				\resizebox{\columnwidth}{!}{%
					\begin{tabular}{@{}c|cc|cc|cc|cc|cc|cc|cc|c@{}}
						\toprule & \multicolumn{12}{c|}{Accuracy [$\%$] per decision window length [s]} &\multicolumn{2}{c|}{Abs. error} & \\ 
						\cmidrule(lr){2-13} & \multicolumn{2}{c|}{60} & \multicolumn{2}{c|}{30} & \multicolumn{2}{c|}{20} & \multicolumn{2}{c|}{10} & \multicolumn{2}{c|}{5} & \multicolumn{2}{c|}{1} &&&\\
						\multirow{-3}{*}{Part.} & true & pred & true & pred  & true & pred & true & pred & true & pred & true & pred & mean & std & \multirow{-3}{*}{\% in CI} \\ \midrule
						1 & 81.9 &      80.4 &      74.3 &      72.9 &       66.2 &      69.1 &      64.1 &      63.8 &      60.9 &      59.9 &      54.7 &      54.4 & 2.2 & 1.8 & 100.0 \\
						2 & 94.4 &      95.3 &      93.8 &      88.4 &    85.6 &      83.6 &      77.8 &      75.6 &      71.5 &      68.8 &      59.7 &      58.7 & 2.6 & 2.1 & 80.0 \\
						3 & 94.4 &      96.0 &      91.7 &      89.5 &     86.6 &      84.8 &      78.0 &      76.7 &      72.8 &      69.7 &      59.8 &      59.1 & 2.4 & 1.6 & 90.0 \\
						4 & 93.1 &      95.4 &      86.1 &      88.4 &     81.9 &      83.5 &      74.8 &      75.5 &      68.8 &      68.7 &      59.4 &      58.6 & 1.7 & 1.2 & 96.7 \\
						5 & 94.4 &      97.4 &      92.4 &      91.7 &     88.9 &      87.1 &      80.6 &      78.9 &      70.6 &      71.6 &      59.8 &      60.1 & 2.0 & 1.6 & 83.3 \\
						6 & 93.1 &      91.2 &      82.6 &      83.2 &    80.1 &      78.4 &      70.6 &      71.1 &      66.7 &      65.3 &      57.4 &      57.0 & 1.6 & 1.3 & 100.0 \\
						7 & 93.1 &      93.5 &      83.3 &      85.9 &      79.2 &      81.0 &      75.5 &      73.3 &      68.2 &      67.0 &      58.1 &      57.8 & 1.7 & 1.4 & 100.0 \\
						8 & 79.2 &      75.7 &      71.5 &      69.0 &      67.1 &      65.7 &      63.0 &      61.3 &      58.6 &      58.0 &      55.2 &      53.6 & 2.4 & 2.6 & 91.7 \\
						9 & 76.4 &      79.3 &      68.1 &      72.0 &     64.4 &      68.4 &      62.3 &      63.3 &      58.2 &      59.5 &      54.1 &      54.3 & 3.5 & 2.9 & 88.3 \\
						10 & 91.7 &      94.3 &      84.7 &      87.0 &     81.5 &      82.1 &      74.5 &      74.3 &      66.4 &      67.8 &      58.8 &      58.2 & 2.0 & 1.4 & 100.0 \\
						11 & 86.1 &      86.5 &      80.6 &      78.3 &     72.7 &      73.9 &      69.0 &      67.5 &      63.2 &      62.6 &      56.6 &      55.7 & 2.0 & 1.6 & 96.7 \\
						12 & 79.2 &      81.8 &      69.4 &      74.1 &    70.4 &      70.1 &      65.0 &      64.6 &      63.0 &      60.4 &      56.3 &      54.7 & 2.7 & 2.2 & 91.7 \\
						13 & 90.3 &      90.9 &      85.4 &      82.8 &     79.6 &      78.1 &      71.5 &      70.9 &      65.2 &      65.1 &      57.2 &      56.9 & 1.8 & 1.6 & 98.3 \\
						14 & 97.2 &      98.7 &      95.1 &      94.3 &    91.2 &      90.1 &      82.9 &      82.0 &      76.5 &      74.1 &      62.4 &      61.4 & 1.5 & 1.1 & 91.7 \\
						15 & 91.7 &      93.1 &      85.4 &      85.4 &     81.9 &      80.5 &      74.3 &      72.9 &      67.4 &      66.7 &      58.7 &      57.6 & 1.7 & 1.1 & 100.0 \\
						16 & 95.8 &      95.4 &      87.5 &      88.4 &     84.3 &      83.6 &      77.8 &      75.6 &      68.1 &      68.9 &      58.8 &      58.7 & 1.6 & 1.5 & 95.0 \\
						\midrule
						\multicolumn{1}{r|}{Mean} & 95.8 &95.4 &87.5 &88.4 &84.3 &83.6 &77.8 &75.6 &68.1 &68.9 &58.8 &58.7 & 2.1 & 1.7 & 94.0 \\
						\bottomrule
					\end{tabular}%
				}
				\label{tab:ind-perf}
			\end{table}
			
			\subsection{Validation on VLAAI network on Fuglsang-2018 dataset}
			\label{sec:validation-vlaai}
			\noindent
			We validate the performance modeling method on the second experimental setup with the VLAAI deep neural network on the Fuglsang-2018 dataset (Section~\ref{sec:vlaai-fuglsang}). Informed by the previous results, we use a $\SI{25}{\second}$ baseline decision window length (closest to $\SI{20}{\second}$ in the range of window lengths in Section~\ref{sec:vlaai-fuglsang}) and $\SI{30}{\minute}$ of estimation data, again with ten repetitions.
			
\begin{figure}
				\centering
				\includegraphics[width=0.8\linewidth]{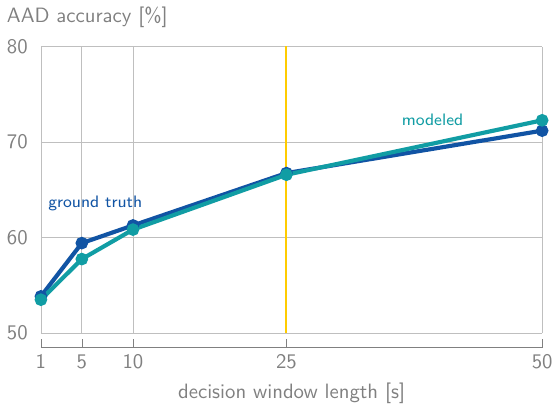}
				\caption{The modeled vs. ground-truth average performance curves of the VLAAI deep neural network on the Fuglsang-2018 dataset, starting from $\SI{25}{\second}$ baseline window length and using $\SI{30}{\minute}$ of estimation data.}
				\label{fig:dataset-wide-vlaai}
			\end{figure}

			Similarly to Fig.~\ref{fig:dataset-wide-wl}, Fig.~\ref{fig:dataset-wide-vlaai} shows the modeled average performance curve across participants vs. the ground-truth one. Table~\ref{tab:ind-perf-vlaai} shows the per-participant mean and standard deviation of the estimation errors across target decision window lengths and repetitions. Both demonstrate comparable performance to the first setup on the Das-2016 dataset, with a mean absolute error across participants of $2.5$ pp (standard deviation $2.1$ pp), and $92.2\%$ coverage by the estimated $95\%$-confidence interval. 
			
			\begin{table}[]
				\setlength{\tabcolsep}{1pt}
				\centering
				\caption{The per-participant prediction results for the VLAAI neural network on the Fuglsang-2018 dataset, starting from $\SI{25}{\second}$ baseline window length and using $\SI{30}{\minute}$ of estimation data.}
					\resizebox{\columnwidth}{!}{%
						\begin{tabular}{@{}c|cccccccccccccccccc|c@{}}
							\toprule & \multicolumn{18}{c|}{Participant} & \\ 
							\cmidrule(lr){2-19} & 1&2&3&4&5&6&7&8&9&10&11&12&13&14&15&16&17&18&Mean\\
							\midrule
							mean abs err. & 2.3 &2.9 &2.5 &2.4 &3.1 &2.4 &2.2 &2.3 &2.4 &1.9 &2.7 &2.7 &2.6 &2.4 &2.1 &2.9 &2.6 &2.5 &2.5   \\
							std abs err. & 1.8 &2.7 &2.8 &1.9 &2.2 &2.4 &1.7 &1.4 &2.3 &1.5 &2.6 &3.0 &2.1 &1.8 &1.6 &2.1 &1.7 &2.4  & 2.1\\
							\% in $95\%$-CI & 88 &94 &92 &98 &86 &92 &94.&90 &94 &96 &98 &98 &90 &88 &96 &88 &82 &96 &92.2 \\
							\bottomrule
						\end{tabular}%
					}
					\label{tab:ind-perf-vlaai}
				\end{table}
				
				\section{Conclusion}
				\label{sec:conclusion}
				\noindent
				We proposed a new AAD accuracy prediction method for correlation-based AAD algorithms that models the entire accuracy-decision window length performance curve starting from the (un)attended correlations at a single window length. Validation across two distinct AAD implementations shows consistent low prediction errors around $2$ percent points. Current limitations include the assumption of two competing speech signals and the requirement for \emph{labeled} baseline correlations. Future work will address multi-speaker scenarios and integration with unsupervised accuracy estimation~\cite{gordo2025unsupervised}.
				
				The proposed AAD accuracy prediction method enables efficient performance curve modeling without continuously re-evaluating AAD performance across multiple decision window lengths, facilitating practical neuro-steered hearing device applications by, e.g., allowing to steer AAD-based gain control systems and providing a theoretical model to translate target AAD accuracies to correlation targets and vice versa.
				
				\bibliographystyle{ieeetran}
				\bibliography{bib-abrev}
				
			\end{document}